\documentclass[aps,pre,twocolumn,groupedaddress,showpacs,showkeys,floatfix]{revtex4-1}
\usepackage{amssymb}
\usepackage[dvipdfm]{graphicx, color}
\usepackage{colordvi}
\usepackage{color}
\begin{document}
\title{The Baldwin effect under multi-peaked fitness landscapes: \\Phenotypic
fluctuation accelerates evolutionary rate}
\author{Nen Saito}
\email[]{saito@complex.c.u-tokyo.ac.jp, Tel.: +81-3-5454-6732}
\author{Shuji Ishihara}
\author{Kunihiko Kaneko}
\affiliation{
Graduate School of Arts and Sciences
The University of Tokyo 3-8-1 Komaba, Meguro-ku Tokyo 153-8902, Japan}
\begin{abstract}
Phenotypic fluctuations and plasticity can generally affect
the course of evolution, a process known as the Baldwin effect.
Several studies have recast this effect and claimed that phenotypic plasticity accelerates
evolutionary rate (the Baldwin expediting effect); however, the validity of this claim is still controversial. 
In this study, we investigate the evolutionary population dynamics of a quantitative genetic
model under a multi-peaked fitness landscape, in order to evaluate the validity of the effect.
We provide analytical expressions for 
the evolutionary rate and average population fitness.
Our results indicate that under a multi-peaked fitness landscape,
phenotypic fluctuation always accelerates evolutionary rate, but
it decreases the average fitness. As an extreme case of the
trade-off between the rate of evolution and average fitness, phenotypic fluctuation is shown to accelerate the error
catastrophe, in which a population fails to sustain a high-fitness  peak.  In the context of our findings, we discuss the role of phenotypic plasticity in 
adaptive evolution.
\end{abstract}
\maketitle

\section{Introduction}
Phenotypic plasticity is a common process, whereby phenotypic changes occur during an organism's life cycle.
In some cases, phenotypic plasticity can emerge as a response to
environmental variation.
For example, the desert locust transitions from a 
	solitary to a gregarious phase, depending on population density~\cite{tawfik1999identification,gilbert2009ecological}.
	In addition, a phenotype can fluctuate randomly, for
	example, as a result of the stochasticity of gene expression, exemplified by recent studies in microorganisms~\cite{spudich1976non,elowitz2002stochastic,sato2003relation,yomo2006responses} and animal
development~\cite{west2003developmental,wernet2006stochastic,feinberg2010stochastic}. Such processes are also interpreted as 
phenotypic plasticity, and are ubiquitous in nature. Given this, the role of random phenotypic plasticity in adaptive evolution warrants further consideration.

Do phenotypic changes in individuals have 
any impacts across generations, and thus, influence their evolution?
At first glance, phenotypic
	plasticity does not seem to affect evolutionary
	processes because only genotypes, rather than phenotypes, are heritable.
Non-heritable phenotypes acquired during the life cycle, however, can affect evolution through natural selection
because the natural selection acts on acquired phenotypes. This does not mean Lamarckism --  heritability of acquired phenotypes,  but indicates that genetically determined plasticity, i.e., an ability to change phenotypes, can affect the evolutionary process.
	This process is referred to as the Baldwin effect, which was first introduced by Baldwin~\cite{baldwin1896new},
	and later recast by Simpson~\cite{simpson1953baldwin}.
		Several studies~\cite{slatkin1976niche,moran1992evolutionary,via1993adaptive,ancel1999quantitative,feinberg2010stochastic,rivoire2011value,frank2011natural2} have revealed that this effect is advantageous under a fluctuating
	environment, as phenotypic plasticity can allow organisms to
	survive sudden environmental changes, and therefore, facilitate adaptation to new environments.
	However, whether phenotypic plasticity is advantageous
	throughout evolution, even under fixed environments, is still elusive.

 Can phenotypic plasticity accelerate evolutionary rate?
This question stems from a recent incarnation of the Baldwin effect known as the "Baldwin expediting
	effect"~\cite{ancel2000undermining}.
Indeed, the seminal work by Hinton and Nowlan~\cite{hinton1987learning} showed
that phenotypic plasticity accelerates evolutionary rate
	even in fixed environments, although, the validity of these findings is
	controversial. For example, several subsequent studies have shown that
	phenotypic plasticity can decelerate evolutionary rate~\cite{anderson1995learning,ancel2000undermining,dopazo2001model},
whereas others have claimed that it accelerates evolutionary rate~\cite{hinton1987learning,gruau1993adding,fontanari1990effect,nolfi1999learning,price2003role,borenstein2006effect,suzuki2007repeated,paenke2007influence}.
		However, it is important to note that the studies\cite{anderson1995learning,ancel2000undermining} in which it was concluded that 
	phenotypic plasticity leads to a deceleration of evolutionary rate dealt
	with uni-modal Gaussian fitness landscapes~\cite{f_Baldwin_1}.
Recently, some studies pointed out that, under a multi-peaked
	fitness, phenotypic plasticity can smoothen
	fitness valleys, and thus, the escape from a local maximum can 
	be enhanced, leading to the acceleration of evolution~\cite{frank2011natural,borenstein2006effect,asselmeyer1996smoothing,lande1980genetic}.  In fact, the acceleration of evolution under a multi-peaked fitness landscape has been confirmed numerically~\cite{nolfi1999learning,gruau1993adding,suzuki2007repeated}.

	The first analytical result addressing evolution under
	a multi-peaked fitness landscape in a fixed environment was provided by Borenstein et
	al.~\cite{borenstein2006effect},
	in which they showed that phenotypic plasticity accelerates 
	evolution.
	Instead of a mutation-selection process, they considered a single random walker
	on a multi-peaked fitness landscape, to mimic evolutionary population
	dynamics in genotypic space; therefore, their
	study did not distinguish between
	the behaviors of a population and an individual.
Thus, in their random walker model the following three points were left unaddressed; first, average population fitness has not been obtained analytically; second, behaviors of the distribution of a population cannot be discussed in their model; and third, the relationship between the acceleration of evolution and average population fitness is unclear.
These points should be clarified in order to fully comprehend the role of
	phenotypic fluctuation (plasticity) in evolutionary population
	dynamics.

In this study, we investigate evolutionary population dynamics, rather than
the single random walker model, under multi-peaked landscapes in a fixed environment,
and study the role of phenotypic fluctuation on the evolutionary rate.
We provide analytical results for the relationship between phenotypic fluctuation, evolutionary rate, and average population fitness
by considering the distribution of genotypes in a population.
	First, we assess whether a population located at a fitness peak can remain around the peak over generations to maintain
	high fitness.
	We show that for both large phenotypic fluctuations and high mutation rates,
	populations fail to keep descendants near the peak over multiple generations, and therefore, they are unable to maintain a high average fitness.
	Second, we evaluate the rate at which a population moves between fitness
	peaks and derive an analytical expression representing the dependency of 
	evolutionary rate on phenotypic fluctuation.
	The obtained result indicates that phenotypic fluctuation
	always enhances diffusion in genotypic space, and thus, it
	accelerates evolution under multi-peaked fitness landscapes.
	
	One consequence of our results is that for a large phenotypic fluctuation, the failure of a
	population to concentrate near a fitness peak results in a decrease in average
	fitness. This can be regarded as ``error
	catastrophe.''
	The concept of error catastrophe refers to the
	collapse of a population that maintains high fitness because of 
	a high mutation rate~\cite{eigen1978hypercycle,eigen1992steps}.
	Our results demonstrate that error
	catastrophe occurs not only because of a high mutation rate, but also
	because of large phenotypic fluctuations. Furthermore, our results illustrate 
	an apparent relationship between error catastrophe and the Baldwin effect, in that 
	error catastrophe is explained as an extreme case of a trade-off between
	an acceleration of evolutionary rate and a decrease in fitness.
	These findings reveal novel aspects of the role of phenotypic plasticity in evolution.

	The organization of this paper is as follows. In
	Sec.~\ref{sec:model1}, we explain how phenotypic fluctuation
	modifies fitness landscapes.
 	In Sec.~\ref{sec:model2}, we
	introduce our model of evolutionary population dynamics with
	phenotypic fluctuation and derive an equation for the time
	evolution of a population distribution in genotypic space.
	In Sec~\ref{sec:result1} and
	Sec~\ref{sec:result2}, we introduce a simple periodic fitness
	landscape as a multi-peaked fitness landscape.
  	In Sec.~\ref{sec:result1}, we
	investigate whether a population can keep its descendants around a peak of fitness, and
	can sustain high average fitness.
	In Sec.~\ref{sec:result2}, we investigate how phenotype
	fluctuation affects evolutionary rate by using
	a transformation of the model equation into the
	Schr$\ddot{\mbox{o}}$dinger equation and adopting WKB approximation.
	In Sec.~\ref{sec:result3}, we investigate a quasi-periodic
	fitness landscape.
	In Sec.~\ref{sec:discussion}, we include concluding remarks.

\section{Model}
\subsection{ Modulated fitness by phenotypic fluctuation}\label{sec:model1}

Let us begin with a discussion of how fitness landscapes are modified by
phenotypic fluctuation (also see~\cite{frank2011natural,anderson1995learning,ancel2000undermining}).
Note that the phenotypic fluctuation here is sometimes referred to the environmental variance in the context of population
genetics and it may also be regarded as an example of phenotypic plasticity.
Here, we analyze asexual population dynamics of a single-locus quantitative genetics model. Suppose that
an individual organism has two types of a one-dimensional quantitative trait:
a heritable trait of genotype $g$ and a non-heritable trait of phenotype $x$.
Genotype $g$ is inherited from its parent, whereas phenotype $x$ is given
	through a genotype-phenotype mapping function $p(x|g;\Sigma)$
	with fluctuation magnitude $\Sigma$.
 	Fitness is assigned to each
	phenotype $x$ and thus is expressed as $F(x)$, given that natural
	selection acts on phenotypes $x$ rather than genotypes $g$. By
	calculating the average with respect to $x$, the effective fitness assigned to
	the genotype $g$ is given as the following equation:

\begin{equation}
	 f(g)=\int _{\Omega} F(x)p(x|g;\Sigma)dx, \label{eq:fitness1}
\end{equation}
where $\Omega$ is the domain of phenotype value $x$.
For simplicity, $p(x|g;\Sigma)$ is assumed to be a Gaussian distribution
with mean $g$ and variance $\Sigma$, and $\Omega$ is assumed to be $\Omega = (-\infty, \infty)$;
\begin{equation}
 p(x|g;\Sigma)=\frac{1}{\sqrt{2\pi
  \Sigma}}e^{-\frac{(x-g)^2}{2\Sigma}}. \label{eq:gpmap}
\end{equation}
From this simplification, Eq.~\ref{eq:fitness1} leads to
\begin{equation}
 f(g)=\int _{\Omega} \frac{F(x)}{\sqrt{2\pi \Sigma}}e^{-\frac{(x-g)^2}{2\Sigma}} dx. \label{eq:fitness2}
\end{equation}
The above equation reveals that phenotypic fluctuations make
a fitness landscape flatter, and its peaks, lower.
Examples
of such modulation effects are shown in Fig.~1(a) and
Fig.~1(b):
By flattening the fitness landscape, fitness peak values are decreased, and thus it acts as a cost of phenotypic fluctuation.

\subsection{The Discrete Time Model and The Approximated Continuous Time
  Equation}\label{sec:model2}
Once $F(x)$ is given, the effective fitness assigned to individual organisms
with genotype $g$ is obtained in Eq.~(\ref{eq:fitness2}).
We here consider the evolutionary population dynamics of $N$ individual organisms with non-overlapping generations.
	Suppose that $N^t(g;\Sigma)$ is the frequency distribution of the
  population at $t$-th generation, and each individual can produce
  descendants that survive in the next generation at a rate $f(g)/\langle f \rangle $.
Here, $\langle \cdot \rangle$ represents an average over
the entire population, (e.g., $ \langle f \rangle \equiv \int_{\Omega} 
  N^t(g;\Sigma)f(g;\Sigma)dg/ \int N^t(g;\Sigma)dg$).
Parental organisms are removed from the population for the next generation.
Due to mutations, genotypes of $t+1$-th generation $g'$ deviate from
  the mother's genotype $g$ as $g'=g+\xi$, where $\xi$ is a random number
  drawn from a Gaussian distribution with a mean zero and variance $D_g$.
This $D_{g}$ corresponds to mutation rate.
This effect of mutations on the distribution $N(g)$ can be written using
the mutation operator $M_{g,g'}$ as $M_{g,g'}[N(g')]=\int_{\Omega} N(g')
  \frac{1}{\sqrt{2\pi D_g}}e^{-(g-g')^2/2D_g} dg'$.
By combining both effects of natural selection and mutations, whole evolutionary dynamics is expressed as
\begin{equation}
 N^{t+1}(g;\Sigma) =M_{g,g'}[ \frac{f(g)N^{t}(g)}{\langle f \rangle } ]\ .\label{eq:model_d}
\end{equation}
With an assumption of sufficiently small $D_g$, this discrete time equation (\ref{eq:model_d}) is approximated by a continuous time equation governed by non-linear Fokker-Planck-like equation:~\cite{f_Baldwin_2} 
\begin{equation}
 \begin{array}{ll}
 \frac{\partial N(t,g;\Sigma)}{\partial t} &=
 \frac{f(g;\Sigma)-\langle f \rangle  }{\langle f \rangle }
 N(t,g;\Sigma)
 +\frac{D_g}{2\langle f \rangle }
  \frac{\partial^2}{\partial g^2}  f(g;\Sigma)N(t,g;\Sigma)\label{eq:model1}.
\end{array}
\end{equation}
In the above equation, $N^t(g;\Sigma)$ is replaced by $N(t,g;\Sigma)$,
which is a function of continuous time.
From this approximation, we can obtain time evolution of a set 
of moment of $N(t,g;\Sigma)$; for instance
\begin{equation}
 \frac{\partial \langle g \rangle}{\partial
  t}=\frac{\langle gf \rangle  -  \langle g \rangle \langle f \rangle
  }{ \langle f \rangle} \label{eq:moment_g}
\end{equation}
\begin{equation}
 \frac{\partial   \langle g^2 \rangle }{\partial
  t}= \frac{  \langle g^2f \rangle -  \langle g^2 \rangle  \langle f \rangle }{ \langle f \rangle}+D_g.\label{eq:moment_g2}
\end{equation}

\section{Results}
\subsection{Localized-extended transition in a periodic fitness landscape}\label{sec:result1}
Here we investigate the model under a simple
multi-peaked landscape,
namely, a periodic one given as $F(x)=1+\cos \alpha x$.
Hence, the effective fitness landscape is expressed as $f(g)=1+\exp (-\alpha
^2 \Sigma /2)\cos \alpha g$ from Eq.~(\ref{eq:fitness2}).
First, we investigate whether the population initially located at a peak of
fitness can stay around the peak
over generations to maintain high fitness.
To examine this, we assume that at the initial state $t=0$ all individuals
in a population are located at $g=0$, and that the genotype distribution
can be approximated by a Gaussian distribution with a mean value $G(t)$
($=\langle g \rangle $) and
variance $V_g(t)$ ($=\langle g^2 \rangle-\langle g \rangle ^2 $) at any generation.
This simplification allows us to calculate the time evolution of
$N(t,g;\Sigma)$ from Eq.~(\ref{eq:moment_g}) and
Eq.~(\ref{eq:moment_g2}), resulting in
\begin{equation}
\hspace{-1cm}
 \begin{array}{ll}
 \frac{dG}{dt}&= -\frac{\alpha V_{g} \sin(\alpha G)e^{-\alpha^2
 (\Sigma+V_{g})/2 }}{ 1+  e^{-\alpha^2(\Sigma+V_g)/2}\cos\alpha g}, \\
 \frac{d V_{g}}{dt}&=-\frac{\alpha^2  V_{g}^2 \cos(\alpha
  G)e^{-\alpha^2 (\Sigma+V_{g})/2 }}{1+  e^{-\alpha^2(\Sigma+V_g)/2}\cos\alpha
 G}+D_g.\\
\end{array}
\label{eq:moment_cos}
\end{equation}
The average population fitness is thus obtained as
\begin{equation}
\langle f \rangle  =\langle 1+ \cos(\alpha
 g)e^{-\frac{\alpha^2\Sigma}{2}} \rangle  =1+\cos(\alpha
 G)e^{-\frac{\alpha^2(\Sigma+V_{g})}{2}}. \label{eq:meanfit}
\end{equation}
At the initial state $t=0$, we assume $G(0)=0$ and $V_g(0)=0$.
Thus, Eq.~(\ref{eq:moment_cos}) leads to $dG/dt=0$ (i.e., $G(t)=0$) and
\begin{equation}
 \frac{d V_{g}}{dt}=-\frac{\alpha^2  V_{g}^2 e^{-\alpha^2
  (\Sigma+V_{g})/2 }}{1+  e^{-\alpha^2(\Sigma+V_g)/2}}+D_g.
\label{eq:moment2}
\end{equation}
The above equation has a fixed point solution when
\begin{equation}
 D_g < \frac{\alpha^2 (V_{g}^{*})^2 e^{-\alpha^2 (\Sigma+V_{g}^{*})/2 }}{1+
  e^{-\alpha^2(\Sigma+V_g ^{*})/2}}, \label{eq:trans2}
\end{equation}
where $V_{g}^*$ gives the minimum of $dV_g/dt$ as given by
\begin{equation}
 V_g ^*=\frac{2\left( 2+W(2e^{-2-\alpha^2\Sigma/2})\right)}{\alpha^2}. \label{eq:vgstar}
\end{equation}
Here, $W(\cdot)$ is the Lambert W-function defined by the inverse function
of $f(W)=W e^W$.
A finite $V_g$ at equilibrium indicates that the  population is
localized around one of the peaks of fitness.
When the inequality (\ref{eq:trans2}) is not satisfied, $V_g$ in Eq.~(\ref{eq:moment2}) always increases in
	time, and thus, the variance of $N(t,g;\Sigma)$ goes to infinity,
	leading to a broad distribution of $N(t,g;\Sigma)$ in genotype space.
This qualitative change in the variance $V_g$ can be referred to as the localized-extended transition.

Simpler representation of the condition Eq.~(\ref{eq:trans2}) is
obtained by using the
approximation $1+\exp[-\frac{\alpha^2(\Sigma+V_g)}{2}] \simeq 1$, which is reasonable around a transition point
because variance $V_g$ is sufficiently large near the transition point.
It yields the condition for a finite variance at equilibrium, namely,
the condition for the localized state as follows:
\begin{equation}
	D_g < \frac{16}{\alpha^2} e^{-\alpha^2 \Sigma/2-2 }.\label{eq:trans3}
\end{equation}
To examine the validity of these arguments, we compare Eq.~(\ref{eq:trans2}) and
Eq.~(\ref{eq:trans3}) with an individual-based simulation (i.e., direct numerical simulation of Eq.~(\ref{eq:model_d})).
Figure 2 shows that both the estimates
agree rather well with the results from the individual-based simulation.

It should be noted that the Gaussian approximation fails when the population is separated
into two subgroups and localized at different peaks of fitness.
Because of this failure, the variance of population estimated
in the individual-based simulation does not coincide with the approximated
value $V_g^{eq}$ calculated
as a solution of Eq.~(\ref{eq:moment2}).
To remedy this failure, we take account of only one peak with the
largest subpopulation, as shown in Appendix \ref{sec:A2}, in which the
procedure to determine the boundary between the localized and extended
phases from the individual-based simulation is also described.

The average population fitness is theoretically estimated as follows:
by solving $dV_g/dt=0$ in Eq.~(\ref{eq:moment2}) by using a quasi-Newton
method, we compute $V_{g} ^{eq}$, and then the average population fitness is
obtained by Eq.~(\ref{eq:meanfit}).
The estimated average fitness is shown in Fig.~3.
We also estimate the average population fitness by using an alternative
approximation method, a harmonic potential approximation, which is 
given below (see Eq.~(\ref{eq:R2}) and also
Eq.~(\ref{eq:har_pot_app}) in Appendix \ref{sec:A4}).
These results are shown in Fig.~3,
and agree rather well with the results obtained from the individual-based simulation.

Both approximated estimations and results from the individual-based
simulation show that in the extended phase, the average population fitness takes a 
value that is as low as that without selection pressure, namely, $\langle f \rangle
=1$. Such a drop in a high fitness value can be interpreted as
``error catastrophe'', i.e., the population fails to
sustain high fitness due to high mutation rate or large phenotypic
fluctuations.
In contrast,
the population in the localized phase is concentrated around the peak
position of the fitness, and thus
can sustain a high average fitness.

With the present analysis based on the Gaussian approximation, the population dynamics in relatively short time
scale are evaluated to examine whether the population is concentrated around the peak. For a longer time scale, population can jump through the fitness valleys, even in the localized phase as we show in the next section.

\subsection{Escape rate analysis in periodic fitness landscape}\label{sec:result2}

Next we focus on the population dynamics of the localized phase.
The population shows diffusion in genetic space, and maintains a high fitness
value for longer time scale.
Figure 4(a) shows examples of time series of mean genotype value $[G]$
obtained in the individual-based simulations, whereas the time course of its
mean square displacement $[G^2]$ is plotted in Fig.~4(b).
In the figure, $[G^2]$ increases in proportion to $t^{1}$, indicating that even in the
localized phase the genotype in population performs random walks by jumping across
a fitness valley.
Through such random
walks, a population can search for novel genotypes with a
higher fitness value in genotype space without suffering from the error
catastrophe.
Thus, the diffusion coefficient of this random walk gives a measure for
the rate of evolutionary innovations, in other words,
evolvability.
It should be noted that the deterministic diffusion in genotype space is expected to be observed for infinite population size, although the results from the individual based simulation show stochastic behavior due to finite population size effect.

The diffusion process in genotype
space constitutes a sequence of jumps across fitness valleys.
As Fig.~1(b) shows, phenotypic fluctuation tends to make
fitness valleys shallower, and thus enhances the jump probability from a peak
to a neighboring peak.
Here, we analytically estimate the jump probability per unit time
$\Gamma$ defined as $\int^{\pi/\alpha}_{-\pi/\alpha}N(t,g; \Sigma)dg \sim e^{-2\Gamma t}$, i.e.,
\begin{equation}\label{eq:defGm}
\Gamma =-\frac{1}{2t}\log \int^{\pi/\alpha}_{-\pi/\alpha}N(t,g; \Sigma)dg,
\end{equation}
with initial condition $N(t=0,g=0;\Sigma)=1$.
Note that the jump probability from a peak of fitness to a neighboring peak per unit time is one half of the escape rate in which the jumps to both neighboring peaks have to be considered.

With a separation ansatz $N(t,g;\Sigma)= e^{\lambda t}\phi(g;\Sigma)$, Eq.~(\ref{eq:model1}) becomes the following eigenvalue problem:
\begin{equation}
 \lambda \phi(g) =
 \frac{f(g)-\langle f \rangle }{\langle f \rangle }\phi(g)+\frac{D}{\langle f \rangle }
  \frac{\partial^2}{\partial g^2}  f(g)\phi(g)\label{eq:model1e},
\end{equation}
where $\phi(g;\Sigma)$ is rewritten as $\phi(g)$ and
$D$ is defined as $D=D_g/2$.
By interpreting $\langle f \rangle$ as an
external control parameter $R=\langle f \rangle$,
we obtain
\begin{equation}
 \Lambda \phi(g) =(f(g)-R )\phi(g)+D
  \frac{\partial^2}{\partial g^2}  f(g)\phi(g)\label{eq:model1e2},
\end{equation}
where $\Lambda$ is given as
\begin{equation}
 \Lambda = R\lambda. \label{eq:lambda}
\end{equation}

After appropriate transformations of variables~\cite{f_Baldwin_3},
 Eq~(\ref{eq:model1e2}) is transformed into
\begin{equation}
\left(-V(y)+d\frac{\partial^2
}{\partial y^2} \right)\psi=\Lambda \psi, \label{eq:sch1}
\end{equation}
where
\begin{equation}\label{eq:potential}
 V(y)= -\left(f(g(y))-R+ \frac{U''(y)}{2} - \frac{(U'(y))^2}{4d}  \right).
\end{equation}
If we use $t_{sc}=-i\hbar t$ and $m_{sc}=\hbar ^2/2d$, this agrees with the familiar form
of Schr$\ddot{\mbox{o}}$dinger equation.

The Schr$\ddot{\mbox{o}}$dinger form Eq.~(\ref{eq:sch1}) with WKB approximation~\cite{simmonds1998first,landauquantum}
allows us to estimate the jump probability between a
peak of fitness and another peak as an escape rate
from a double well potential,
assuming that the probability distribution is located only in the left well at $t=0$, as is illustrated in
Fig.~5.
Details of the derivation are given in Appendix \ref{sec:A4}.

Now, we consider the same fitness landscape in Sec.~\ref{sec:result1}: $f(g)=1+\chi\cos\alpha g$ and
$\chi=\exp (-\alpha^2 \Sigma /2)$, where $\chi$ decreases as phenotypic fluctuation $\Sigma$ increases.
Using quasi equilibrium approximation (see details in Appendix \ref{sec:A4}),
the average population fitness $R$ for this landscape can be derived as follows:
\begin{equation}
 R= 1+\chi-D\frac{\alpha^2}{4}\chi-\sqrt{\frac{\alpha^2
  D(1+\chi)\chi}{2}+\frac{\alpha^4 D^2 \chi^2}{16}}. \label{eq:R2}
\end{equation}
This evaluation of average fitness $R$ at the quasi equilibrium
agrees rather well with the results from the individual-based simulation, as is shown in
Fig.~3 (the thick line represents the evaluation in
Eq.~(\ref{eq:R2}) ), indicating that the quasi equilibrium approximation is valid.

Using WKB approximation and $R$ estimated above, we can evaluate the jump probability per unit time $\Gamma$ as
\begin{equation}
\begin{array}{ll}
\Gamma&=\frac{\sqrt{2\alpha^2D\chi(1+\chi)} }{R\pi}
\exp \left( -\frac{2}{\sqrt{D}}
\int
						  _{b} ^{\pi/\alpha} \sqrt{
							      \frac{\chi(1-\cos\alpha
							      g)+\Theta (g)}{1+\chi\cos\alpha
							      g}}
							       dg
					 \right), \label{eq:er2}
\end{array}					 
\end{equation}
where
\begin{equation}
\begin{array}{ll}
 \Theta(y)&=-\sqrt{\frac{D}{2}(1+\chi)\alpha^2\chi+\frac{D^2}{16}\alpha^4\chi^2}\\
  &+\frac{D\chi^2\alpha^2}{8}\frac{\sin^2 \alpha g}{1+\chi\cos \alpha
  x}-\frac{\alpha^2 \chi D}{4}(1-\cos \alpha g)
\end{array}
\end{equation}
Note that $g=b$ is a point where the integrand becomes zero i.e., $b$ satisfies
\begin{equation}
 \chi(1-\cos\alpha b)+\Theta (b)=0
\end{equation}
The effective diffusion coefficient $D_e$ is given by
\begin{equation}
 D_e =2\frac{\Gamma}{2}\left(\frac{2\pi}{\alpha} \right)^2.\label{eq:ed1}
\end{equation}
The factor $2$ in the above equation appears since the escape
rate from a peak of fitness to the left neighbor peak has to be
incorporated, as well as that to the right neighbor peak.
Equations (\ref{eq:er2}) and Eq.~(\ref{eq:ed1})
describe the relation between phenotypic fluctuation $\Sigma$ and
evolvability $D_e$.

To examine the validity of the estimation of the escape rate $2\Gamma$ in
Eq~(\ref{eq:er2}), we compare Eq.~(\ref{eq:er2}) with that obtained
numerically by the individual-based simulation of
Eq.~(\ref{eq:model_d}) (details of the estimation of the escape rate
from the individual-based simulation are explained in Appendix \ref{sec:A3}).
As shown in Fig.~6, the two agree rather well.
These results shown in Fig.~6 indicate
that phenotypic fluctuation always accelerates the evolutionary
rate, while large phenotype fluctuation decreases the average population fitness, as is shown in Fig.~3.
Therefore, the evolvability and fitness
values are in a trade-off relationship.

\begin{figure}
\includegraphics[width=9cm]{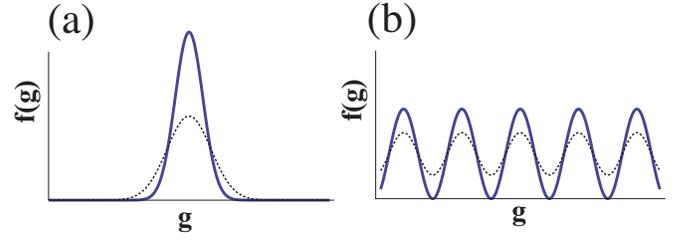}
\caption{(COLOR ONLINE)
Fitness landscapes. (a) A unimodal fitness landscape with (dotted line) and without (line) phenotypic fluctuation.
Phenotypic fluctuation modulates an original
fitness landscape to that with a lower peak and a wider
range.
(b) A multi-modal fitness landscape with (line) and without (dotted line)
phenotypic fluctuation.
}
\end{figure}

\begin{figure}
\includegraphics[width=6cm]{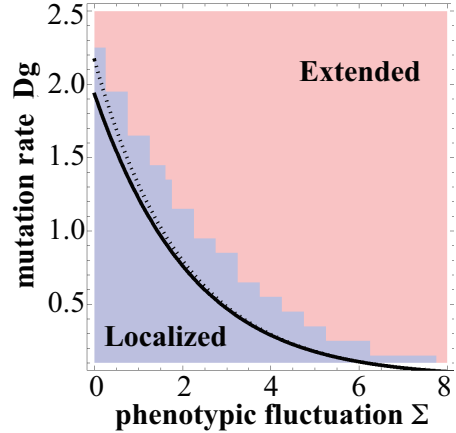}
\caption{(COLOR ONLINE)
A phase diagram for the localized-extended transition.
The black line indicates the phase boundary estimated in Eq.~(\ref{eq:trans2})
and the dotted line indicates the boundary estimated in
Eq.~(\ref{eq:trans3}). The boundary between the blue and red regions
is estimated from the individual-based simulation. Details of the estimation are described in Appendix \ref{sec:A2}. 
}
\end{figure}

\begin{figure}
\includegraphics[width=6cm]{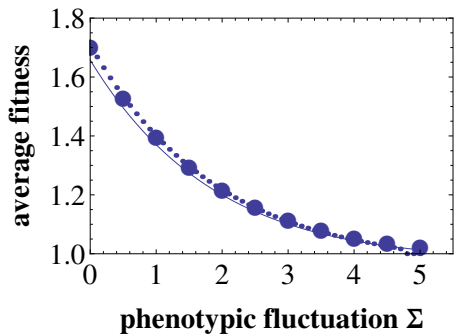}
\caption{(COLOR ONLINE)
Phenotypic fluctuation decreases average population fitness.
The fitness are plotted against the phenotype fluctuation
$\Sigma$ for $D_g =0.2$ and $\alpha = 1$. 
Circles indicate results from the individual-based simulation.
The dashed line indicates the approximated value obtained from
Eq.~(\ref{eq:meanfit}), where $V_g ^{eq}$ is calculated numerically from
Eq.~(\ref{eq:moment2}). Thick line indicates the approximated value
obtained from a harmonic potential approximation in Eq.~(\ref{eq:R2}).
Eq.~(\ref{eq:trans3}) indicates that the error catastrophe occurs as $\Sigma > 4.76$
for $\alpha =1.0$, as $\Sigma > 2.80$.
Because $\alpha$ can be always set as unity by choosing appropriate
scaling of $x$,
results at $\alpha =1$ is shown.
}
\end{figure}

\begin{figure}
\includegraphics[width=9cm]{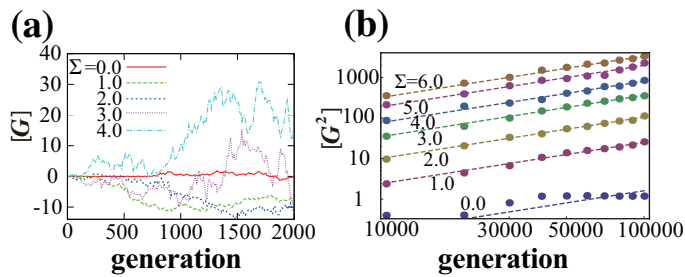}
\caption{(COLOR ONLINE)
Time series of the distribution of population in genotype space.
(a) A time series of G (average g over population at each time)
obtained by the individual-based simulation.
(b) A log-log plot of the time series of $[G^2 ]$, where $[]$
indicates the sample average.
In both cases, the average $[\cdot ]$ is computed over 100 samples.
$[G^2]$ for each $\Sigma$ can be fitted to a line with a slope equal to
one. This indicates that $G$ performs random walks.
In both figures, $N=500$ and $D_g=0.3$ are used.
}
\end{figure}

\begin{figure}
\includegraphics[width=4cm]{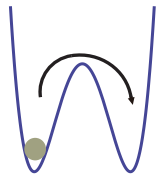}
\caption{(COLOR ONLINE)
A double well potential.
The probability is localized at the left
well in the initial condition. A jump probability among peaks of fitness can be
interpreted as the escape rate from the left well of the double well potential.
}
\end{figure}

\begin{figure}
\includegraphics[width=6cm]{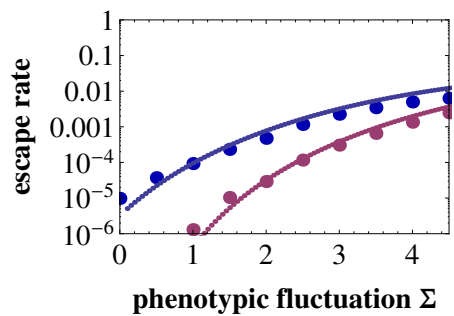}
\caption{(COLOR ONLINE)
Escape rate from a peak of fitness.
The escape rate $2\Gamma$ for periodic potential with respect to phenotype fluctuation $\Sigma$. 
Circles indicate
results from the individual-based simulations. Dashed
lines indicate the values estimated by using the WKB approximation in
Eq.~(\ref{eq:er2}).
The upper blue line and circles are results for $F(g)=2+\cos \alpha g$ and the lower red 
 line and circles are results for $F(g)=1+\cos \alpha g$.
Parameters $N=500$ and $D_g=0.2$ are used.
Because $\alpha$ can be always set as unity by choosing appropriate
scaling of $x$,
results at $\alpha =1$ is shown.
}
\end{figure}

\subsection{Mixed cosine potential }\label{sec:result3}
To demonstrate that our analysis is not limited to the simple
periodic potential,
we consider the quasi-periodic fitness landscape;
$F(x)= A_1(1+\cos \alpha_1 x )+A_2(1+\cos \alpha_2 x )$.
From Eq.~(\ref{eq:fitness1}), the effective fitness landscape can be calculated as
\begin{equation}
f(g,\Sigma)=A_1(1+e^{-\frac{\alpha_1^2\Sigma}{2}} \cos \alpha_1 g) +A_2(1+e^{-\frac{\alpha_2^2\Sigma}{2}} \cos \alpha_2 g) .\label{eq:mixed_cos}
\end{equation}
An example of the effective fitness $f(g)$ is illustrated in Fig.~7.

Using Gaussian approximation of $N(g;\Sigma)$, we obtain the time evolution of mean value $G$ and variance $V_g$ of $N(g;\Sigma)$
as
\begin{equation}
 \begin{array}{ll}
 \frac{dG}{dt}&=\\
 & -V_g \frac{ A_1 \alpha_1  \sin(\alpha_1
			   G)e^{-\alpha_1^2 (\Sigma+V_{g})/2 }+ A_2 \alpha_2  \sin(\alpha_2
			   G)e^{-\alpha_2^2 (\Sigma+V_{g})/2 }}{A_1(1+  e^{-\alpha_1^2(\Sigma+V_g)/2}\cos\alpha_1
 G )+A_2(1+  e^{-\alpha_2^2(\Sigma+V_g)/2}\cos\alpha_2
 G )} \\
 \frac{d V_{g}}{dt}&=\\
 &-V_g ^2\frac{A_1\alpha_1^2  \cos(\alpha_1
  G)e^{-\alpha_1^2 (\Sigma+V_{g})/2 }+A_2\alpha_2^2 \cos(\alpha_2
  G)e^{-\alpha_2^2 (\Sigma+V_{g})/2 }}{A_1(1+  e^{-\alpha_1^2(\Sigma+V_g)/2}\cos\alpha_1
 G)+A_2(1+  e^{-\alpha_2^2(\Sigma+V_g)/2}\cos\alpha_2
 G)}\\
 &+D_g \label{eq:moment3}
\end{array}
\end{equation}

Nullcline analysis of Eq.~(\ref{eq:moment3}) is given in
Fig.~8(a) - (c), against the three
values of $\Sigma$:
for small $\Sigma$ (Fig.~8(a)), a population can be
localized at around each local
maximum of fitness; for middle value of $\Sigma$ (Fig.~8(b)),
a population can be localized only at around a larger local maximum of fitness; for
large $\Sigma$ (Fig.~8(c)), all fixed points
disappear, and $V_g$ diverges, which indicates the extended phase.
These results suggest the existence of two transitions:
the localized-localized
transition and the localized-extended transition.

\subsubsection*{(i) Localized-localized transition }
For the simplicity of discussion, we assume $\alpha_1>\alpha_2$ in the following argument.
Here, we define the
localized-localized transition as the transition where a population
cannot be
localized at around the lowest local maximum of $V(x)$.
Thus, we investigate whether a population can be localized around $G_{lm}$, which satisfies
\begin{equation}
 \cos(\alpha_1 G_{lm})=1, \label{eq:coscon1}
\end{equation}
\begin{equation}
  \cos(\alpha_2 G_{lm})=-1 \label{eq:coscon2}.
\end{equation}
In other words, we determine the conditions on which $V_g$ has finite value for $t \to
\infty$ for a population at $G=G_{lm}$ from the following equation:
 \begin{equation}
 \begin{array}{ll}
 \frac{d V_{g}}{dt}&=\\
 &-V_g ^2\frac{A_1\alpha_1^2  e^{-\alpha_1^2 (\Sigma+V_{g})/2 }-A_2\alpha_2^2  e^{-\alpha_2^2 (\Sigma+V_{g})/2 }}{A_1(1+  e^{-\alpha_1^2(\Sigma+V_g)/2})+A_2(1- e^{-\alpha_2^2(\Sigma+V_g)/2})}\\
 &+D_g\label{eq:moment3_2}.
\end{array}
\end{equation}
Solving the above equation with an initial condition $V_g(0)=0$
by numerical integration, we determine the transition point where $V_g$ diverges. In
Fig.~9, the blue line shows the localized-localized
transition point obtained by the numerical integration.

Intuitive estimation is provided by assuming that the critical
value of $V_g$ is $\tilde{V_g} \sim \left(
\frac{\pi}{2\alpha_1}\right)^2$, which corresponds to the square of the width of
a smaller peak.
From this assumption, the transition point can be evaluated by
\begin{equation}
\begin{array}{ll}
D_g &\le (\frac{\pi}{2\alpha_1}) ^4\\
&\cdot \frac{A_1\alpha_1^2  e^{-\alpha_1^2 (\Sigma+(\frac{\pi}{2\alpha_1})^2)/2 }-A_2\alpha_2^2  e^{-\alpha_2^2 (\Sigma+(\frac{\pi}{2\alpha_1})^2)/2 }}{A_1(1+  e^{-\alpha_1^2(\Sigma+(\frac{\pi}{2\alpha_1})^2)/2})+A_2(1- e^{-\alpha_2^2(\Sigma+(\frac{\pi}{2\alpha_1})^2)/2})} \label{eq:moment4}.
\end{array}
\end{equation}
The above estimate agrees well with the results of the numerical
integration of Eq.~(\ref{eq:moment3_2}) (see Fig.~9).

\subsubsection*{(ii) Localized-extended transition}
To investigate the localized-extended transition,
we consider the behaviors of the population distribution located at the global maximum
$G=0$.
We assume that, near the transition point, $V_g$ is large enough to consider
$e^{-\alpha_1^2(\Sigma+V_g)/2}$ as negligible, but small enough to
interpret $e^{-\alpha_2^2(\Sigma+V_g)/2}$ as a finite value.
From this assumption, Eq.~(\ref{eq:moment3}) can be rewritten as
\begin{equation}
\hspace{-1cm}
 \begin{array}{ll}
 \frac{d V_{g}}{dt}&=-V_g ^2\frac{A_2\alpha_2 ^2  e^{-\alpha_2 ^2
  (\Sigma+V_{g})/2 }}{A_2(1+  e^{-\alpha_2 ^2(\Sigma+V_g)/2})+A_1}+D_g \label{eq:moment5}.
\end{array}
\end{equation}
Here, the same procedures to derive Eq.~(\ref{eq:trans2}) and Eq.~(\ref{eq:trans3})
are applied. In Fig.~9, we show the estimated
localized-extended transition point.

\begin{figure}
\includegraphics[width=6cm]{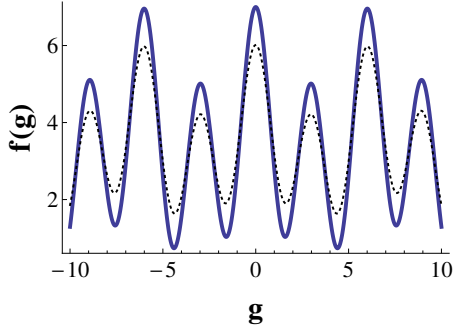}
\caption{(COLOR ONLINE)
A mixed cosine fitness landscape.
The landscpe is given by
Eq.~(\ref{eq:mixed_cos}), with (dotted line) and without (line)
phenotypic fluctuation. Parameters $A_1=2.5, A_2=1.0, \alpha_1=2.1, \alpha_2=1$ are used.
}
\end{figure}

\begin{figure}
\includegraphics[width=9cm]{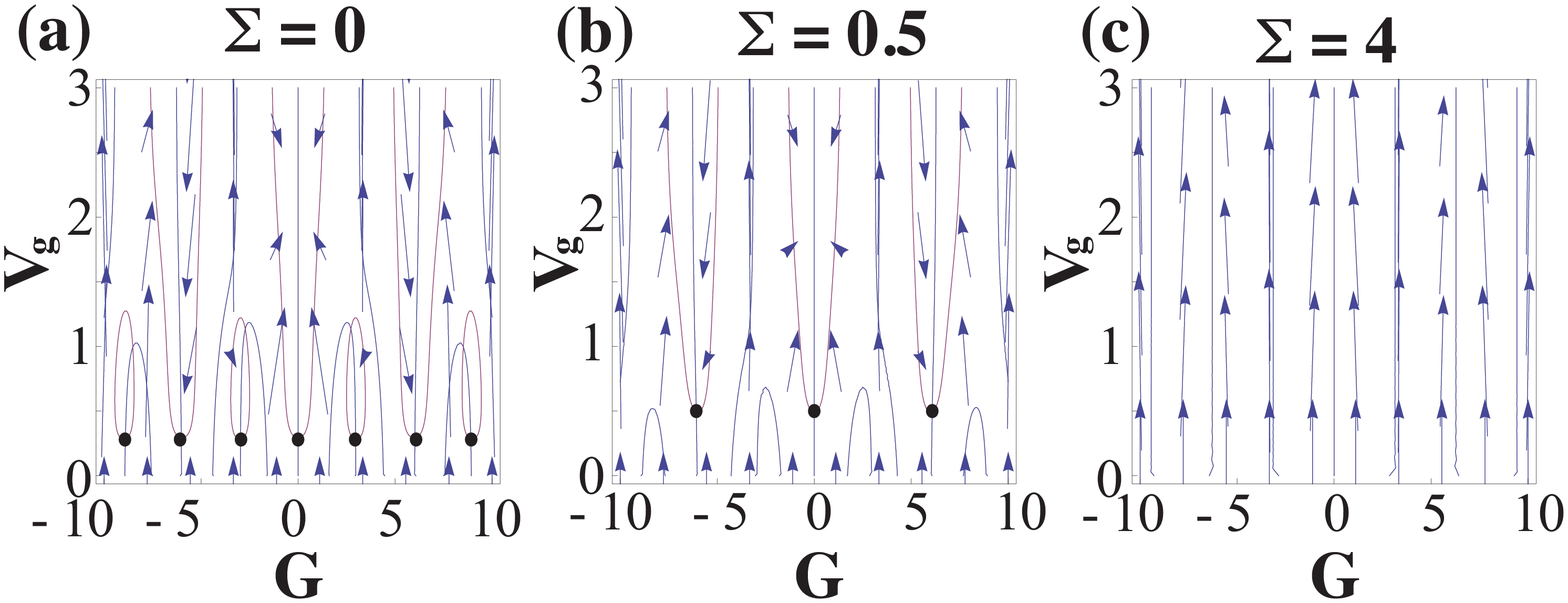}
\caption{(COLOR ONLINE)
Nullcline analysis for Eq.~(\ref{eq:moment3}).
$A_1=2.5, A_2=1.0,
\alpha_1=2.1, \alpha_2=1$ and $D_g=0,1$ are used. The horizontal axis
represents $G$, and the vertical axis, $V_g$. Black dots
indicate stable fixed points. (a) for small $\Sigma$ ($\Sigma=0$), stable fixed
points appear around each local maximum of fitness.
(b) For $\Sigma=0.5$, some of the stable fixed
points vanish, and thus, there are only fixed points around
$G=\frac{2\pi}{\alpha_2}\times n$, $n=(0, \pm 1,\pm 2,...)$.
This indicates that a population is localized only around a peak associated
with $\alpha_2$. (c) For large $\Sigma$ ($\Sigma=4$), all fixed points
vanish. In this case, the population goes to the extended phase.
}
\end{figure}

\begin{figure}
\includegraphics[width=9cm]{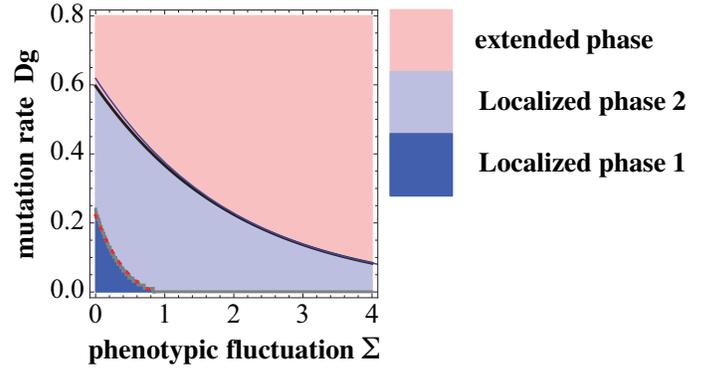}
\caption{(COLOR ONLINE)
A phase diagram of the localized-extended and 
localized-localized transition for a mixed cosine fitness landscape.
Parameters are set as $A_1=2.5$, $A_2=1.0$, $\alpha_1=2.1$, $\alpha_2=1$, and $D_g=0.1$.
The boundary between the localized phases 1 (the left lower phase) and 2 (the middle phase) is
estimated by numerical integration of Eq.~(\ref{eq:moment3_2}) (gray line)
and by Eq.~(\ref{eq:moment4}) (red dotted line). The boundary between the localized phase 2
and the extended phase (the right upper phase) is estimated by $D=\frac{16}{\alpha_2 ^2}\frac{A_2}{A_1+A_2}e^{-\alpha_2 ^2\Sigma/2 -2}$.
}
\end{figure}

\subsection{General Rugged Landscape}
Next, we confirm our scenario that phenotypic fluctuation accelerates evolution for a general rugged fitness landscape numerically.
Here, we consider a landscape $ F(x)=1+\sum_{k}^{M}\frac{\alpha_{k}}{\sum_{j}^{M}\alpha_{j}}\cos\left( \frac{2\pi k}{L}x+\frac{2\pi L}{k}\phi_{k}\right)$, where $\alpha_{k}$ and $\phi_{k}$ are uniform random number ranging over $[0,1]$.
Although this landscape is periodic with period $L$, it can be regarded as a random landscape within length $L$.
From Eq.~(\ref{eq:fitness1}), the effective fitness landscape can be calculated as
\begin{equation}\label{eq:rugfit}
f(g,\Sigma)=1+\sum_{k}^{M}\frac{\alpha_{k}}{\sum_{j}^{M}\alpha_{j}} e^{-2\pi^{2}k^{2}\Sigma/L^{2}} \cos\left( \frac{2\pi k}{L}x+\frac{2\pi L}{k}\phi_{k}\right).
\end{equation}
Figure 10-(a) shows $f(g,\Sigma)$ for $M=10$, $L=20$. 
Using this effective fitness landscape, we perform the individual-based simulation. 
As shown in Fig.~10-(b), the time course of mean square displacement, $[G^{2}]$, of the mean genotypic value $G$ of each generation indicates that diffusion constant in genotype space increases monotonically with $\Sigma$.
This demonstrates that phenotypic fluctuation enhances the evolutionary rate (For some case of $F(x)$, the diffusion shows non-monotonic change against $\Sigma$ for small values of $\Sigma$ (data not shown), but for large $\Sigma$, it shows monotonic increase with $\Sigma$).

\begin{figure}
\includegraphics[width=9cm]{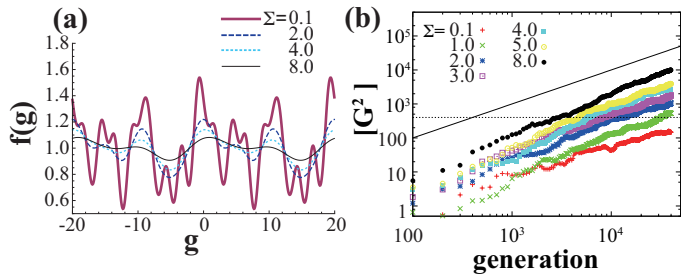}
\caption{(COLOR ONLINE)
(a) A general rugged fitness landscape in Eq.~(\ref{eq:rugfit}) for $M=10$ and $L=20$.
Each line represents $f(g,\Sigma)$ for $\Sigma=0.1, 2.0, 4.0$ and $8.0$.
(b) Time course of mean square displacement, $[G^{2}]$, of the mean genotypic value $G$ of each generation for the fitness landscape in (a).
The graph is plotted in log-log scale.  
Average $[\cdot$] is taken over 100 trials.  Parameters $N=500$ and $D_{g}=1.0$ are used.
Black thick line shows $[G^{2}]=generation$ for reference, while the black dotted line indicates the point where $[G^{2}]=L^{2}=400$; under this line, the landscape can be regarded as a non-periodic random potential.}
\end{figure}

\section{Discussion}\label{sec:discussion}
In the present study, the evolutionary population dynamics of a simple quantitative genetics model
under a multi-peaked fitness landscape is investigated by focusing on the role of phenotypic fluctuation on
the evolutionary process.
The main results of this study are summarized in the following points (i)-(iv):

(i) The relationship between average fitness and
phenotypic fluctuation. Analytical expression is obtained either by Gaussian
approximation of a population distribution or by a harmonic potential approximation.
By the Gaussian approximation, average fitness is given
by Eq.~(\ref{eq:meanfit}) with a solution of ODE Eq.~(\ref{eq:moment2})
that requires numerical computation, while by the harmonic potential approximation,
the average fitness is analytically estimated as Eq.~(\ref{eq:R2}) (or
Eq.~(\ref{eq:har_pot_app})).
As is shown in Fig.~3,
both approximations agree rather well with the results from the individual-based
simulation of evolutionary population dynamics described by Eq.~(\ref{eq:model_d})
and indicate that average population fitness decreases as
the magnitude of phenotypic fluctuation $\Sigma$ increases.

(ii) Error catastrophe due to large phenotypic fluctuation.
Because of either a high mutation rate or large
phenotypic fluctuation, a population fails to keep its descendants around a peak of fitness, and thus is unable to sustain a high-fitness phenotype.
Such a decrease in fitness is known as ``error catastrophe.''
Although the error catastrophe usually refers to the failure in maintaining
the population due to a high mutation rate, our result demonstrates that the error
	catastrophe can also occur as a result of phenotypic fluctuation.
In our model, phenotypic fluctuation always enhances the risk of
error catastrophe, whereas an interesting exception to this has been reported in which phenotypic fluctuation can suppress
the error catastrophe under a step-like fitness landscape in a high-dimensional
genotype space~\cite{sato2007evolution}.

(iii) Dependence of evolutionary rate on the magnitude of phenotypic
fluctuation.
We provide an analytical expression of the rate at which a population jumps
from one peak to another peak in the multi-peaked fitness landscape, which is 
expressed as the evolutionary rate.
The obtained expression Eq. (\ref{eq:er2}) shows that phenotypic plasticity
generally enhances evolutionary rate. 
Such an acceleration has been noted previously~\cite{gruau1993adding,frank2011natural,nolfi1999learning,borenstein2006effect,suzuki2007repeated}; however, our results provide the first analytical expression including both the effects of natural selection and mutation.
We also confirm numerically that the phenotypic fluctuation accelerates evolution for a general rugged fitness landscape, and thus the present conclusion is not limited to a periodic fitness landscape but is expected to be valid for general landscapes, such as Fig.~10-(a).

Our claim of the acceleration of evolution by phenotypic fluctuation is different
from Fisher's fundamental theorem~\cite{fisher1930genetical}, which claims that genotypic
variance is proportional to evolutionary rate.
In Fisher's theorem, the phenotypic plasticity or fluctuations are
not considered, and the effect of mutation is also ignored, whereas in our results,
genetic variance, phenotypic effect, and the effect of mutation are incorporated.
A possible relationship between the phenotypic fluctuation and genotypic variance was also discussed in~\cite{sato2003relation}.

(iv) Phenotypic fluctuation exhibits trade-off relationships with the
acceleration of evolution and degradation of fitness.
This conclusion is reached by combining the results (i), (ii), and (iii).
This trade-off relationship is illustrated by both
Fig.~3 and Fig.~6, in which greater phenotypic
fluctuation results in a smaller fitness value and a larger evolutionary rate.
In addition, the transition to the extended phase (the error catastrophe)
for larger phenotypic fluctuation can be interpreted as an extreme situation of this
trade-off.
It is well established that
an increase in mutation rate leads to the same trade-off
relationship between the evolutionary rate and degradation of fitness,
and that error catastrophe occurs as an extreme case of the trade-off.
Our results indicate that phenotypic fluctuation
introduces an additional mutation rate, although
the origin of the phenotypic fluctuation is essentially different from
that of mutation.
Note that this ``additional mutation rate'' effect of
phenotypic fluctuation works only in case of multi-peaked fitness
landscape,
but not in case of a unimodal fitness landscape~\cite{anderson1995learning,ancel2000undermining,dopazo2001model}.

Throughout the present study, we interpret 
	random phenotype fluctuation as a kind of phenotypic plasticity
	(i.e., the phenotypic distribution of a single genotype is independent of
	environment and fitness (see Eq.~(\ref{eq:gpmap}))).
	Several previous studies addressing the Baldwin effect also adopted
	random fluctuation as phenotypic
	plasticity~\cite{frank2011natural,anderson1995learning,ancel2000undermining,borenstein2006effect}.
	Our results indicate that even random phenotypic fluctuation accelerates the evolutionary rate under a multi-peaked
	fitness landscape.
	Another interpretation of the phenotypic plasticity is
	that phenotype distribution of a given genotype depends on the fitness landscape, referred to as responsive plasticity~\cite{hinton1987learning,fontanari1990effect,gruau1993adding,price2003role,borenstein2006effect,suzuki2007repeated}.
	The difference between phenotypic fluctuation
	and responsive plasticity is only in the modulation of a fitness landscape by phenotypic changes, and therefore,
	the techniques that we adopted in the present study (e.g., the transformation of Fokker-Planck-like
equation into the Schr$\ddot{\mbox{o}}$dinger equation, WKB
	approximation) are applicable after a
	modulated fitness is obtained; these techniques are generally useful for investigations of evolutionary dynamics~\cite{PhysRevE.67.061118,asselmeyer1997evolutionary,sato2006distribution,sato2007evolution}.

In our model, we assume that both genotype and phenotype are represented as a one-dimensional continuous value and the mapping between them is straightforward (assumptions often employed in quantitative genetics~\cite{falconer1981introduction,anderson1995learning,ancel2000undermining}). Because such simplicity could miss potential importance of high dimensionality and complexity in real genotype-phenotype mapping, we give brief remarks here. A high dimensional binary genotype model with a simple fitness landscape was adopted in some old studies~\cite{hinton1987learning,fontanari1990effect}, especially the work by Hinton and Nowlan~\cite{hinton1987learning} demonstrated the validity of Baldwin effect in the model.  Recently, relevance of phenotypic fluctuation (noise in developmental process) to mutational robustness has been studied by models with high-dimensional genotype space and with non-trivial genotype-phenotype mapping~\cite{kaneko2007evolution,sakata2009funnel,sakata2011replica}.  Although these studies did not intend to validate the Baldwin effect, some of their results share a mechanism of the acceleration effect of evolution; their conclusion suggested that smoothening fitness landscape by phenotypic fluctuation prevents from trapping in local maximum of fitness, which is also important for the Baldwin effect.
For this reason, it is reasonable to expect that the Baldwin effect is valid even for the case with high dimensional genotype space under a rugged fitness landscape. To confirm this statement, however, further studies are required.

In this paper we study an infinite population model in which population is always genetically polymorphic and there is no fixation event of a single genotype.
It will be important for future research to extend our study to include cases with finite populations, in which both the genetic drift and
phenotypic fluctuations have to be taken into account seriously.

So far, mounting experimental evidence supports that fitness landscapes
in real organisms are
multi-peaked~\cite{fong2005parallel,korona1994evidence,hayashi2006experimental}.
In addition, it has also been
shown that stochasticity and random fluctuations in phenotypes are
common in living
systems~\cite{spudich1976non,elowitz2002stochastic,sato2003relation,yomo2006responses}.  
In the presence of such
fluctuation and multi-peaked fitness landscape, our analytical study has
demonstrated the relationship between phenotypic fluctuation and
evolutionary rate, as well as the relationship between the Baldwin
effect and error catastrophe. Therefore, our findings have broad
implications for the study of evolutionary dynamics.

\section*{Acknowledgment}
This work was partially
supported by a Grant-in-Aid for Scientific Research (No. 21120004)
on Innovative Areas ``Neural creativity for communication'' (No.
4103) and the Platform for Dynamic Approaches to Living System from
MEXT, Japan.

\bibliographystyle{unsrt}

\begin{thebibliography}{10}
\bibitem{f_Baldwin_3}
By adopting variable as $y(g)=\sqrt{\frac{d}{D}}\int^g
  \frac{1}{\sqrt{f(g')}} dg'$ and
  $\psi(y)=e^{U(y)/2d}\sqrt{\frac{f(g(y))}{d}}\phi(g(y))$, where
  $U'=\frac{\partial U}{\partial y}=\sqrt{dD}\frac{\partial f/ \partial
  g}{2\sqrt{f}}$, Eq~(\ref{eq:model1e2}) is transformed into
  Eq.(\ref{eq:sch1}).
\end{thebibliography}

\appendix
\section{Detail of the individual-based simulation}\label{sec:A1}
We use an individual-based evolutionary population dynamics model based on an algorithm similar
to a genetic algorithm (GA);
this simulation is a direct computation of Eq.~(\ref{eq:model_d}).
In this model, each individual in a population has a genotype $g$ and an effective fitness
$f(g)$, which is calculated in Eq.~(\ref{eq:fitness1}). We use
$F(x)=1+\cos(\alpha x)$ in Sec.~\ref{sec:result1} - Sec.~\ref{sec:result2}  and $ F(x)=A_1(1+\cos \alpha_1 x )+A_2(1+\cos
\alpha_2 x )$ in Sec.~\ref{sec:result3}.

In our simulation, total population size is fixed to $N$.
At the first generation, each individual has the same genotype $g=0$. At the
$t$-th generation, $N$ individuals for the next generation are selected from
the current generation; each individual is sampled with probability
$f(g)/N \overline{f }$, where $\overline{f}=\sum_i ^N f(g_i)/N$. Mutations are applied to a selected population
in a way in which $g$ of each selected individual is mutated as $g \to g+\xi$.
$\xi$ is a random number drawn from a Gaussian distribution
$e^{-g^2/2D_g}/\sqrt{2\pi D_g}$, where $D_g$ corresponds to the mutation rate. 
After a selection and mutation procedure, we obtain the $t+1$-th generation.
Iterating the procedure, we simulate the population dynamics.

\section{Determination of the phase boundary in Fig.~2}\label{sec:A2}
To restore the failure of the Gaussian approximation, we take account
of only one interval with an integer $n$, $S_n = \{ g|~ (2\pi n - \pi)/\alpha < g < (2\pi n +
\pi)/\alpha \} $  that has the largest subpopulation. At each time step
in the simulation, variance of the population $V_{sim}$ is evaluated by
using the subpopulation in $S_n$ by the following equation.
\begin{equation}
V_{sim} = \frac{1}{n_S}\sum_{i \in S_n}(g_i- \frac{2\pi n}{\alpha})^2
\end{equation}
Here, $n_S$ indicate the number of individuals in $S_n$. 
We also evaluated $\tilde{V}_g^{\ast}$ by truncated Gaussian distribution as
\begin{equation}
\tilde{V_g} ^{*}= \frac{\int ^{\pi/\alpha} _{-\pi/\alpha} g^2 \exp(- g^2 / 2V_g ^{*}
)/\sqrt{2\pi V_g ^{*}} dg}{\int  ^{\pi/\alpha} _{-\pi/\alpha} \exp(- g^2 / 2V_g ^{*}
)/\sqrt{2\pi V_g ^{*}} dg},
\end{equation}
where $V_g^{\ast}$ is calculated by Eq.~(\ref{eq:vgstar}). 
In Fig.~2, localized
(extended) phase is determined by the condition that the time average
of $V_{sim}$ is smaller (larger) than $\tilde{V}^{*}$.

Note that the above estimate of $V_{sim}$ using only subpopulation in
$S_n$ underestimates the variance of the population, because broader
distributions of coexisting subpopulations in neighboring peaks are
ignored. 
The agreement between the boundaries by theory and simulation in Fig.~2
could be improved by taking appropriately into account a correction for
the estimate of the variance.

\section{Details of the evaluation of the escape rate from the
  individual-based simulation}\label{sec:A3}
To evaluate the escape rate of a population from a peak of fitness,
we perform the individual-based simulation of Eq.(\ref{eq:model_d}) with 
the following boundary condition: at the initial condition $t=0$, all
individuals in a population are located at $g=0$; when an individual
escapes from the region $(-\pi/\alpha,\pi/\alpha)$, $g$ of the
individual is transformed into $g+M\pi/\alpha$, where $M$ is
a sufficiently large integer that the individual never returns to the region $(-\pi/\alpha,\pi/\alpha)$.
Note that this transformation does not change the fitness of each
individual.
Using this boundary condition, we compute time series of the
average fraction of a population in the region $(-\pi/\alpha,\pi/\alpha)$ and
then estimate the decreasing rate of the average fraction as the
escape rate. Here, the average fractions are calculated over 200 trials
for $\alpha = 1.0, 1.2$ and 3000 trials for $\alpha =0.8$.
Then the escape rates are calculated as $2\Gamma$ in Eq.(\ref{eq:defGm}).

\section{WKB approximation}\label{sec:A4}
Here, we assume that the population is located only around a peak of fitness.
This assumption allows us to suppose that the probability distribution of the
Schr$\ddot{\mbox{o}}$dinger equation is only
located in the left well at $t=0$, as is illustrated in Fig.~5.
Within a time-scale much shorter than that of a jump, we can assume that
a population is approximately in an equilibrium state (a quasi equilibrium) in a harmonic potential, which is
derived from the Taylor expansion around a minimum of $V(y)$, $y=y^*$
\begin{equation}
\begin{array}{ll}	
 V(y)&\simeq V(y^*) +\frac{1}{2}V''(y^*)(y-y^*)^2\\
 &=V(y^*)+\frac{1}{2}\left(-\Xi''(y^*)+\frac{(U''(y^*))^2}{2d} \right)(y-y^*)^2,\label{eq:harm_V}
 \end{array}
\end{equation}
where $\Xi (y)=f(y)-R$.
The second term in Eq.~(\ref{eq:harm_V}) is interpreted as $\frac{1}{2}m_{sc}\omega ^2 y^2$, and therefore,
\begin{equation}
 \frac{1}{2}m_{sc}\omega ^2 = \frac{\hbar ^2\omega^2
  }{4d}=\frac{1}{2}\left(-\Xi''(y^*)+\frac{(U''(y^*))^2}{2d} \right) \label{eq:omega},
\end{equation}
where $\omega$ is the angular frequency of the harmonic potential.
The ground energy of the harmonic potential is given by
\begin{equation}
\begin{array}{ll}
 E_0 &=V_{min}+\frac{\hbar \omega}{2}\\
 &=-\left(\Xi(y^*)+\frac{U''(y^*)}{2}
				     \right)+\sqrt{\frac{d}{2}\left(-\Xi''(y^*)+\frac{(U''(y^*))^2}{2d}
						      \right)}.
\end{array}
\end{equation}
At the quasi equilibrium, the eigenvalue $E_0$ should be
zero, yielding
$$
 R=\hat{f}(y^*)+\frac{U''(y^*)}{2}-\sqrt{\frac{d}{2}\left(-\Xi''(y^*)+\frac{(U''(y^*))^2}{2d}
							      \right)}\label{eq:R1}
$$
\begin{equation}
=f(g^*)+\frac{D}{4}\frac{\partial^2 f}{\partial g^2}-\sqrt{\left(\frac{Df(g^*)}{2}|\frac{\partial ^2 f}{\partial g^2}(g^*)|+\frac{D^2}{16}(\frac{\partial ^2 f}{\partial g^2}(g^*))^2
							      \right)},\label{eq:har_pot_app}
\end{equation}
where $\hat{f}(y^*)=f(g(y^*))=f(g^*)$,
$\Xi''(y^*)=\frac{Df(x^*)}{d}\frac{\partial ^2 f}{\partial g^2}|_{g=g^*}$ and
$U''(y^*)=\frac{D}{2}\frac{\partial ^2 f}{\partial g^2}|_{g=g^*}$.

The escape rate of the double well potential can be calculated by the difference
$\Delta \Lambda$
between the smallest eigenvalue $\Lambda_0$ and the second smallest eigenvalue $\Lambda_1$~\cite{PhysRevE.67.061118,risken1996fokker,van2007stochastic}. To derive
this difference, the most successful technique is WKB approximation,
in which sufficiently small and smooth potential $V(y)$
are assumed. By WKB approximation, the escape rate $\Gamma$ of
double well potential can be estimated as~\cite{landauquantum, PhysRevE.67.061118}
\begin{equation}
 \begin{array}{ll}
\Gamma &=\frac{\Delta \Lambda}{2R}=\frac{\hbar\omega }{R\pi}\exp \left(
						  -\frac{2\sqrt{2m}}{\hbar}\int
						  _b ^a \sqrt{
							       V(y)-E_0}
							       dy
				     \right)\\
&=\frac{\hbar\omega }{R\pi}\exp \left(
						  -\frac{2}{\sqrt{d}}\int
						  _b ^a \sqrt{
							       V(y)-E_0} dy \right),\label{eq:er1}
\end{array}
\end{equation}
where $y=b$ is a point in which the integrand becomes zero and $y=a$ is a
point in which the integrand becomes a maximum (i.e., the central peak of
potential in Fig.~5 ).
It should be noted that, from Eq.~(\ref{eq:lambda}), the time scale of
Eq.~(\ref{eq:model1}) is R times slower than Eq.~(\ref{eq:model1e2}).
The integrand in Eq.~(\ref{eq:er1}) can be estimated as
\begin{equation}
 \begin{array}{ll}
 \sqrt{V(y)-E_0} &= \sqrt{-\left(\Xi(y)+ \frac{U''(y)}{2} - \frac{(U'(y))^2}{4d}  \right)}\\
 &= \sqrt{R-\hat{f}(y)-\left(\frac{U''(y)}{2} - \frac{(U'(y))^2}{4d}  \right)}\\
& = \sqrt{f(g^*)-\hat{f}(y) +\Theta(y)  },
\end{array}
\end{equation}
where
\begin{equation}
\begin{array}{ll}
 \Theta(y)&=-\sqrt{-\frac{Df(g^*)}{2} \frac{\partial^2 f(g^*)}{\partial g^2}+\frac{D^2}{16}(\frac{\partial^2 f(g^*)}{\partial g^2})^2
							      }\\
							      & +\frac{D}{8f}(\frac{\partial f(g)}{\partial g})^2+\frac{D}{4}(\frac{\partial^2 f(g^*)}{\partial g^2}-\frac{\partial^2 f(g)}{\partial g^2}).
\end{array}
\end{equation}
Thus, equation Eq.~(\ref{eq:er1}) becomes
\begin{equation}
 \begin{array}{ll}
\Gamma &=\frac{\hbar\omega }{R\pi}\exp \left(
						  -\frac{2}{\sqrt{d}}\int
						  _b ^a \sqrt{
							       f(g(y^*))-f(g(y))
							       +\Theta (y)}
							       dy
				\right)\\
& =\frac{\hbar\omega }{R\pi}\exp \left(
						  -\frac{2}{\sqrt{d}}\int
						  _{g^*}^{g_a} \sqrt{
							       f(g^*)-f(g)+\Theta(y) } \frac{dy}{dg}dg \right)\\
 &=\frac{\sqrt{2Df(g^*)|f''(g^*)|} }{R\pi}\exp \left(
						  -\frac{2}{\sqrt{D}}\int
						  _{g^*} ^{g_a} \sqrt{
							      \frac{f(g^*)-f(g)+\Theta(y(g))}{f(g)}}
							       dg
				\right).
\end{array}
\end{equation}
Note that, in the above equation, we use $\hbar
\omega=\sqrt{D|f''(g^*)|(2f(g^*)+D|f''(g^*)|/4)}\simeq\sqrt{2Df(g^*)|f''(g^*)|}$
from Eq.~(\ref{eq:omega}).

\appendix

\end{document}